# Advanced Receiver Autonomous Integrity Monitoring: Impact of Time-Correlated Pseudorange Measurement Noise

Jindrich Dunik, Martin Orejas, *Honeywell International*

**BIOGRAPHY**

**Jindrich Dunik** is a scientist within Advanced Technology Europe, Honeywell International. He received his MSc degree in automatic control in 2003 and the PhD degree in cybernetics in 2008, both from University of West Bohemia, Czech Republic. Until 2010 he was with the Department of Cybernetics, University of West Bohemia, focusing on state estimation methods. Within Honeywell, he is currently working in areas of inertial and satellite based navigation systems and integrity monitoring methods. He is also author or co-author of more than 30 technical papers (both journal and conference) in the fields of nonlinear filtering and system identification.

**Martin Orejas** is a senior scientist within Advanced Technology Europe at Honeywell International. He received a MSc degree in electronic engineering from National University of Comahue, Argentina, and a double MSc degree in space technology and instrumentation from Lulea UT, Sweden, and space automation and control, from Czech Technical University, Czech Republic. He has been working as a researcher with special focus on satellite based navigation systems, inertial systems, integrated navigation, and advanced integrity monitoring methods. He is currently leading several projects related to the development of future multiconstellation GNSS receivers and GNSS/INS hybrid navigation systems.

**ABSTRACT**

The paper deals with the allocation of the probability of false alert within the advanced receiver integrity monitoring method. Namely, the stress is laid on the correct computation of the probability of false alert per sample under assumption of time-correlated pseudorange noise. Detailed analysis of the dependence of the probability of false alert per sample on the measurement noise time constant is given and a numerical algorithm for the correct computation of the probability is proposed. The algorithm is illustrated using a numerical example.

**I. INTRODUCTION**

GNSS is becoming a cornerstone for airborne navigation systems for all categories of aircraft and the shift to a multi-constellation GNSS scenario is expected to accentuate this trend. This new scenario, which will benefit from the modernization of GPS and GLONASS and deployment of Galileo and COMPASS, will provide a significant increase in the accuracy and integrity of the GNSS based navigation solution. This improved performance is, in turn, anticipated to allow GNSS based navigation systems to support some of the most stringent phases of flight, in particular approaches with vertical guidance down to a minimum decision height of 200 feet. However, in order to fully exploit this improved performance, changes need to be made to the integrity monitoring scheme (mainly because of a higher classification of a failure condition for an integrity fault). Currently, one of the most promising integrity monitoring schemes foreseen for a multi-constellation scenario is the Advanced Receiver Autonomous Integrity Monitoring (ARAIM) developed by the joint EU-US group [1].

ARAIM is the enhanced version of the Solution Separation method [2] where the protection levels are computed on the basis of

- position estimate covariance matrices of the full-solution and all sub-solutions,
- values of unmodeled biases in pseudorange measurements,
- probabilities of false alert and missed detection (or the probability of hazardous misleading information instead) which are allocated among all sub-solutions.

The paper is solely focused on the correct computation of the probability of false alert per sample. The probability per sample can be derived from continuity requirement by dividing it by the total number of samples. Such an approach is correct if the noise in pseudorange measurements is not correlated in time (white). If the noise is time-correlated, then the common approach for computing the probability of false alert per sample is to divide the continuity requirement by the number of independent samples. The number of independent samples is often computed as the time period over which the continuity requirement is specified divided by the time constant of the measurement error. Unfortunately, this common approach is not correct and might yield optimistic values of the probability of false alert per sample.

The goal of the paper is to propose the correct approach how to take into account the time-correlation of the measurement noise in computation of the probability of false alert per sample. The proposed approach is based on the computation of the "correction" coefficient to the probability of false alert per sample computed under assumption of the white measurement noise. The correction coefficient is a function of the probability of false alert per sample and the time constant of the noise correlated in time. The theoretical analysis is presented and general relations for coefficient computation are derived. However, because of not trivial analytical solution to the relations, a numerical algorithm is proposed and verified using numerical illustrations.

The rest of the paper is organized as follows. Sections II and III are focused on the introduction of ARAIM and on discussion regarding standard computation of the probability of false alert per sample, respectively. In Section IV and V, the analytical and numerical relations for correct computation of the probability are discussed and their application in ARAIM is illustrated. Finally, concluding remarks are given in Section VI.

## II. ADVANCED RECEIVER AUTONOMOUS INTEGRITY MONITORING METHOD

ARAIM method is one of the most promising algorithms being proposed to meet stringent requirements imposed by the approach with vertical guidance, e.g. LPV-200 approaches [1]. The method for pure GNSS navigation systems is briefly introduced in the following text.

The method can be understood as an extension of the solution separation method. Thus, it is based on statistical tests of estimates provided by the full-solution (processing all available GNSS measurements) and sub-solutions (processing a subset of the measurements selected according to the specified GNSS fault states to be mitigated).

### A. Snap-shot navigation solution (full-solution and sub-solutions)

Let the pseudorange error vector of all available pseudorange measurements be denoted as $\Delta\rho_0$ [3]. The vector is related to the offset in the receiver position $\Delta x_0$ (subscript "0" denotes the full-solution) by the relation
$$\Delta\rho_0 = H_0 \Delta x_0 + w.$$
The first three components of $\Delta x_0$ are the position correction to the initial position estimate $\hat{x}$ (e.g., in North-East-Down (NED) coordinate system) and the fourth component is the correction to the time bias. Matrix $H_0$ is the measurement matrix based on the line-of-sight (LOS) vectors among the receiver initial position and satellites positions. Pseudorange error vector $\Delta\rho_0$ is computed as the measured pseudoranges minus the ones estimated on the basis of the estimate $\hat{x}$. Variable $w$ represents the measurement noise and is supposed to be a Gaussian zero-mean random variable with the covariance matrix $R_0$, i.e.,
$$w \sim N(0, R_0).$$
The solution to the position and time correction computed according to the (weighted) least-squares method is given by
$$\Delta x_0 = S_0 \Delta\rho_0 = (H_0^T W_0 H_0)^{-1} H_0^T W_0 \Delta\rho_0,$$
where $S_0$ is the solution matrix and $W_0 = R_0^{-1}$ is the weighting (usually diagonal) matrix. The initial position is then updated to get the corrected receiver position estimate according to
$$\hat{x}_0 = \hat{x} + \Delta x_0.$$
The error of the estimated position offset to the actual position offset $\Delta x$ is given by
$$\tilde{x}_0 = \Delta x - \Delta x_0 = \Delta x - S_0 H_0 \Delta x_0 - S_0 w = S_0 w,$$
where the equality $S_0 H_0 = I$ was used. The covariance matrix of the estimation error of the full-solution is equal to
$$P_0 = S_0 R_0 S_0^T = (H_0^T W_0 H_0)^{-1}.$$
The $n$-th sub-solution (not taking the $n$-th pseudorange measurement into account) is computed on the basis of all but the $n$-th pseudorange measurement[1], i.e., according to
$$\Delta x_n = S_n \Delta\rho_0, n = 1, 2, \ldots, N,$$
where the solution matrix is
$$S_n = (H_n^T W_n H_n)^{-1} H_n^T W_n$$
with
$$H_n = E_n H_0,$$
$$W_n = E_n W_0,$$
$$E_n = I - e_n e_n^T.$$
The vector $e_n$ denotes the $n$-th column of the identity matrix of appropriate dimension, i.e., $e_n = I(:,n)$. The solution matrix can be further treated as
$$S_n = (H_0^T E_n W_0 H_0)^{-1} H_0^T E_n W_0.$$
The error of the estimated position offset to the actual position offset $\Delta x$ is analogously given by
$$\tilde{x}_n = \Delta x - \Delta x_n = -S_n w,$$
where the equality $S_n H_0 = I$ was used. The covariance matrix of the estimation error of the sub-solution is then equal to
$$P_n = S_n R_0 S_n^T = (H_0^T E_n W_0 H_0)^{-1}.$$

ARAIM (and also almost any other integrity monitoring method) is based on computing three quantities; test statistics, decision threshold, and protection levels which are introduced below.

---

[1] For the sake of simplicity, $N$ pseudorange measurements and $N$ sub-solutions are assumed (only single failures are considered). Generally, $N$ is rather related to the number of specified fault states.

## B. Test statistics

The test statistic is represented by the separation between the full set and all subset solutions. Each separation constitutes a test statistic. The separation in vertical dimension[2] is computed according to

$$d_{Vn} = (\hat{x}_0(3) - \hat{x}_n(3)), n = 1,2,\ldots,N,$$

where $\hat{x}_0(3)$ is the third component of the vector $\hat{x}_0$ representing position correction in Down direction.

For each of the statistics, the decision threshold is set to ensure that the false alert probability allocated to that test statistic is not exceeded.

## C. Decision threshold

Decision threshold is based on the computed covariance matrix $dP_n$ characterizing separation between the full-solution and $n$-th sub-solution. The separation between the full-solution and the sub-solution is

$$d_n = \hat{x}_0 - \hat{x}_n = \hat{x} + \Delta x_0 - \hat{x} - \Delta x_1 = (S_0 - S_n)w$$

The separation covariance matrix is then equal to

$$dP_n = cov[d_n] = (S_0 - S_n)cov[ww^T](S_0 - S_n)^T$$
$$= (S_0 - S_n)R_0(S_0 - S_n)^T.$$

Based on the probability of false alert per sample $P_{FA}$, number of sub-solutions $N$, and taking into account the contribution of nominal biases in the range measurements, the vertical thresholds are given by[3]

$$D_{Vn} = \sqrt{dP_n(3,3)}Q^{-1}\left(\frac{P_{FA}}{2N}\right) + DB_{Vn}, \forall n,$$

where $dP_n(3,3)$ is the element of matrix $dP_n$ in the third row and column and $Q^{-1}$ is the inversion function to the complement of the one-sided standard normal cumulative distribution function

$$Q(x) = \frac{1}{\sqrt{2\pi}} \int_x^{\infty} e^{-\frac{t^2}{2}} dt.$$

Variable $DB_{Vn}$ takes into account contribution of unmodeled (nominal) biases in pseudorange measurements of which expression can be found e.g., in [1].

The alarm is raised to notify user about the fault if any $d_{Vn}(= d_n(3)) < -D_{Vn}$ or $d_{Vn} > D_{Vn}$.

Note that the separation covariance matrix is computed on the basis of non-integrity assured pseudorange measurement error definition as the alert threshold affects the continuity only.

## D. Vertical protection level

The VPL for all subsolutions is computed as

$$VPL_n = D_{Vn} + a_{Vn},$$

where the error bounds corresponding to the missed alert probability $P_{MD}$ are given by

$$a_{Vn} = \sqrt{P_n(3,3)}Q^{-1}(P_{MD}) + AB_{Vn}$$

and $AB_{Vn}$ is computed using the maximum bias magnitude.

In this case, the covariance matrix is computed on the basis of integrity assured measurement error descriptions.

## III. PROBABILITY OF FALSE ALERT AND MEASUREMENT NOISE PROPERTIES

From the previous section it follows, that specification of the probability of false alert per sample, i.e., $P_{FA}$, is one of the key ARAIM design parameters. In the following text computation of $P_{FA}$ is discussed.

### A. Continuity of navigation information

The total probability of false alert, denoted hereafter as $P_{FA,total}$, can be derived from the continuity requirement.

Continuity is related to the capability of the navigation system to provide a navigation output with the specified accuracy and integrity throughout the intended operation, assuming availability of the information at the start of operation. For example, in case of the LPV-200 approach the continuity in terms of the total probability is equal to $8\times10^{-6}$ per 15*sec* for horizontal and vertical dimension. If the probability is equally split between horizontal and vertical dimension, then $P_{FA,total}$ is equal to $4\times10^{-6}$ per 15*sec* for each dimension [9], [10], [11].

Probability of false alert basically defines the expected number of the false alerts within a given time frame. It means, it defines "how often" might be the absolute value of the separation $d_{Vn}$ greater than the decision threshold $D_{Vn}$. This clearly depends on the properties of the position estimate errors of the full-solution $\tilde{x}_0$ and the $n$-th sub-solution $\tilde{x}_n$ which defines the covariance matrix $dP_n$ and subsequently the threshold.

### B. Position estimates errors and pseudorange measurement noise modeling

In fact, the properties of the position estimate errors are driven by the properties of the pseudorange measurement noise. If the pseudorange noise is uncorrelated in time (i.e., white), then the position estimate error is white as well and vice versa, if the measurement noise time correlated, then position error is also correlated. Moreover, the time constants of the position error and pseudorange measurement noise are the same (as the weighted least-square algorithm used for position solution is a memoryless algorithm processing the GNSS related measurements only and all measurements are assumed to have the noise with the same time constant).

The pseudorange measurement noise is formed by pseudorange smoothing, tropospheric error, multipath error, etc. Each of these contributions might have possibly different time constant, but often the time constant used in models is about 100 seconds [4], [5].

---

[2] Requirements on vertical channel are usually considered to be more stringent than requirements on horizontal one. Therefore, ARAIM has been designed with a focus on vertical dimension only.
[3] It is assumed that the probability of false detection is equally allocated among all test statistics.

One more comment is related to the usage of a hybrid navigation system. From the conceptual point of view, ARAIM can be used also with the hybrid navigation system. Then, instead of multiple computations of the least-squares solutions, the set of hybridization filters (e.g., the extended Kalman filter [6], [3]) needs to be run. However, as the filter integrates the measurements from various sensors (e.g., GNSS, inertial sensors, altimeter), each possibly with a noise with different time constant, the time constant of the position error estimates cannot be generally specified as a function of the GNSS properties only. In this case, an analysis assessing the impact of the particular sensors on the position estimate properties has to be performed.

### C. $P_{FA}$ under assumption of time-uncorrelated noise

Under assumption of the pseudorange measurement white noise, the per-sample probability $P_{FA}$ is computed simply by dividing the continuity requirement by the total number of samples in the given time period. That means, the per-sample probability under assumption of white noise for LPV-200 is

$$P_{FA}^{white} = \frac{4 \times 10^{-6}}{15}.$$

### D. $P_{FA}$ under assumption of time-correlated noise – standard approach

Such an approach, unfortunately, does not hold when the samples (the position estimates) are not independent. In this case, the probability of false alert is not constant for all subsequent samples but is conditioned by the value of the previous sample. This can be explained using an intuitive example as follows: if a time-correlated noise is considered (e.g., modeled by the Gauss-Markov (GM) process) then the values of two subsequent noise samples do not mutually differ "much". The time evolution of the subsequent samples of a GM process is mostly affected by the driving noise, with usually much smaller variance than the variance of the whole process. The subsequent measurements are then also similar and thus, if in one time instant the false alert is not observed, in the subsequent time instant the false alert is "more likely" to be not observed as well. In other words it means the number of observed false alerts is less if the measurement noise is time-correlated than the noise is white. This implies that the probability of false alert per sample cannot be derived from the continuity requirement by simply dividing it by the total number of samples.

A common approach to deal with the time-correlated measurement errors is to compute the probability of false alert per sample by dividing the continuity requirement by the number of independent samples, where the number of independent samples is computed as the time period over which the continuity requirement is specified, e.g., 1 hour or 150 seconds, divided by the time constant of the measurement error.

Indeed, the common approach for LPV-200 under the assumption of the time-correlated noise is to set the per-sample probability as

$$P_{FA}^{corr,common} = 4 \times 10^{-6}.$$

That means, the per-sample probability is set to be equal to an overall budget defined for the continuity [10]. The whole interval for which the continuity is defined is thus considered as "a single sample" due to the time-correlation of the measurement noise[4].

Unfortunately, this approach is not correct and might yield optimistic values for the probability of false alert per sample. In the following part, therefore, a correct approach is proposed.

### IV. COMPUTATION OF PROBABILITY OF FALSE ALERT PER SAMPLE

The proposed approach strictly follows the definition of the continuity [4]:

"The continuity of a system is the ability of the total system (comprising all elements necessary to maintain aircraft position within the defined airspace) **to perform its function without interruption during the intended operation**. More specifically, **continuity is the probability that the specified system performance will be maintained for the duration of a phase of operation**, presuming that the system was available at the beginning of that phase of operation, and predicted to exists throughout the operation."

Respecting the continuity definition (and following the discussion from the previous section), the sample-based probability $P_{FA}$ need not necessarily be constant for all times within the operation duration. In fact, it is necessary to consider the sample-based probability time-variant and conditioned by the past values.

More specifically, the conditional (real) probability in this context stands for the sample-based probability $P_{FA,K}$ under assumption that no false alert has been detected in the previous time instants $k = 0,1,\ldots,K-1$ (test statistic at time $k$, $d_{Vn,k}$, is below the threshold $\forall k$), i.e.,

$$P_{FA,K}^{cond} = prob\big(d_{Vn,K} \notin S_D \big| d_{Vn,0} \in S_D, \ldots, d_{Vn,K-1} \in S_D\big),$$

where $S_D$ is the interval defined by the threshold computed on the basis of $P_{FA}^{white}$ allocated for a given sub-solution. That means, $S_D = \langle -D_{Vn}, D_{Vn} \rangle$ and $P_{FA,K}^{cond} \leq P_{FA,K-1}^{cond}$. The conditional sample-based probability at time $K$, $P_{FA,K}^{cond}$, thus, generally differs from the unconditional one $P_{FA,K}$. The unconditional probability is constant over the whole period and it is equal to the per-sample probability under assumption of white noise..

---

[4] In [10] it is stated "The allowable continuity loss probability per sample was taken to be the same as the probability per 15-second interval at $4 \times 10^{-6}$."

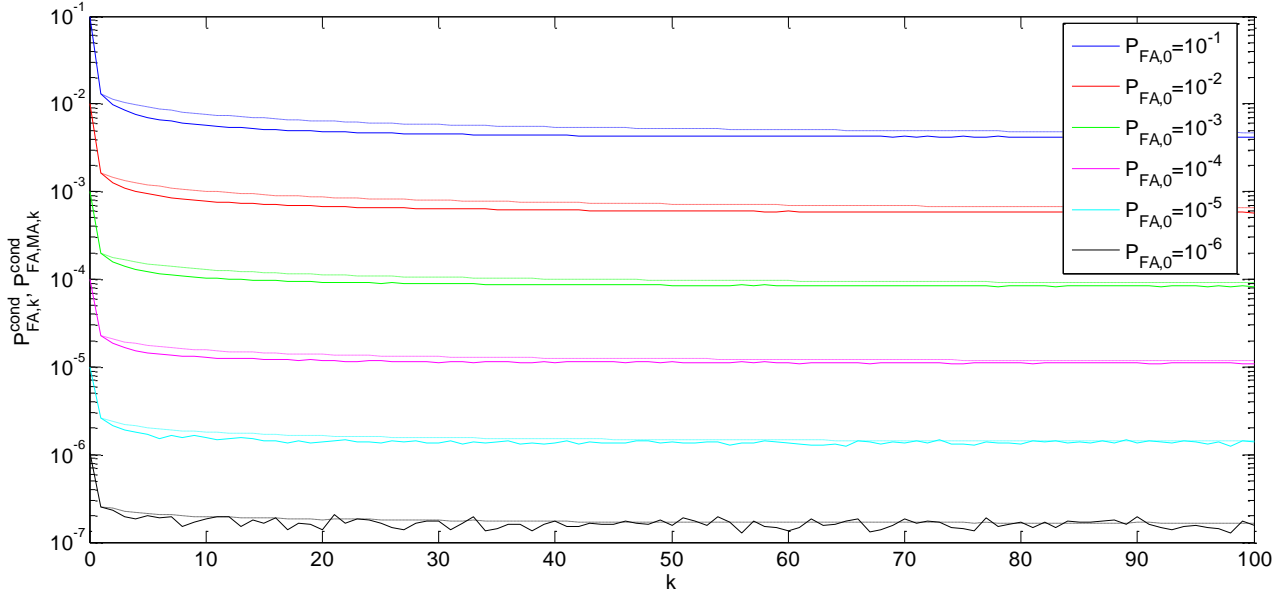

Figure 1: Time behavior of conditional per-sample probability of false alert (conditional $P_{FA}$ - solid line, its MA - dashed line).

Analytical computation of $P_{FA,K}^{cond}$ is given in Appendix A together with a related discussion. Also, as the analytical solution is not trivial, a numerical Monte-Carlo integration based algorithm is proposed in the appendix and tested.

As an example, the time behaviors of $P_{FA,K}^{cond}$ under assumption of time-correlated process with the time constant 100 $sec$ are shown in Figure 1 for different values of $P_{FA,0}$, which can be understood as of $P_{FA}^{white}$ allocated for a subsolution. In the figure, also the moving average

$$P_{FA,MA,K}^{cond} = \frac{1}{K}\sum_{k=1}^{K} P_{FA,k}^{cond}$$

is plotted. The moving average represents the overall conditional per-sample probability estimate.

It can be seen that if $P_{FA,0} = 10^{-6}$, then the conditional probability $P_{FA,100}^{cond}$ is approximately $0.16 \times 10^{-6}$ and $P_{FA,MA,k}^{cond} \approx 0.18 \times 10^{-6}$. Therefore, the (real) conditional per-sample probability of false alert at time $k = 100$ is more than 6 times lower than the probability computed on the assumption of the white measurement noise.

One possibility to get the expected number of false alerts also in case of the time-correlated signals, it is, therefore, necessary to increase $P_{FA}^{white}$ according to

$$P_{FA}^{cond} = P_{FA}^{white} \times c_{corr},$$

where $P_{FA}^{cond}$ is the overall compensated per-sample probability and $c_{corr}$ is the correction coefficient depending on $P_{FA}^{white}$ and time constant of the measurement noise. The correction coefficient might be computed as

$$c_{corr} = \frac{P_{FA}^{white}}{P_{FA,MA,Kend}^{cond}},$$

where the time instant $Kend$ is related to the length of the intended operation.

Another possibility is to use in threshold computation directly the computed conditional (and time-variant) probability $P_{FA,k}^{cond}$ instead of computation of an overall averaged value $P_{FA}^{cond}$.

## V. NUMERICAL EXAMPLE: CORRECT PER-SAMPLE $P_{FA}$ COMPUTATION IN ARAIM

The impact of the considered computations of per-sample probability of false alert on the resulting VPL computed by ARAIM is illustrated in this section.

As was mentioned above, the LPV-200 continuity requirement for vertical dimension is $4 \times 10^{-6}$ per 15$sec$. Assuming white measurement noise, the per-sample probability is

$$P_{FA}^{white} = \frac{4 \times 10^{-6}}{15}.$$

If the measurement noise is correlated in time with the time constant of 100$sec$, the common approach for computation of per-sample probability is [10]

$$P_{FA}^{corr,common} = 4 \times 10^{-6}.$$

However, according to the proposed approach, the correct conditional per-sample probability for time-correlated set-up is

$$P_{FA}^{cond} = P_{FA}^{white} \times c_{corr} = \frac{4 \times 10^{-6}}{15} \times c_{corr},$$

where $c_{corr}$ is the correction coefficient taking into account time-correlation of the measurement noise (and underlying test statistics). The coefficient computed by the proposed algorithm for the given set-up is

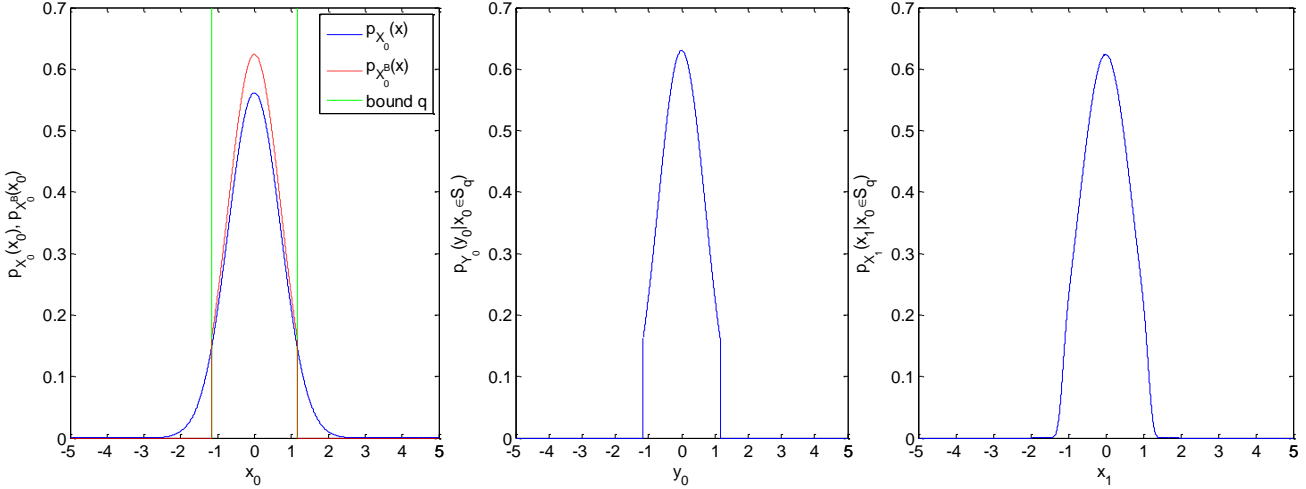

Figure 2: Evolution of conditional probability density function for GM process.

$c_{corr} \approx 5$.

It can be seen, that the common approach leads to the larger allocation of $P_{FA}$ (approx. three times) which subsequently results in slightly optimistic estimates of "Dn-terms" in ARAIM (causing larger number of observed false alerts) and the respective protection levels. The minimum, maximum, and mean VPLs for three above mentioned choices of $P_{FA}$ are given in Table 1. VPLs are computed for dual-frequency and dual-constellation set-up in 2664 different locations and 100 different time instants. Thus, 266,400 different values of VPL were computed for each choice of $P_{FA}$. Budgets of the pseudorange measurements noise (both integrity assured and non-integrity assured) are similar to those used in [1].

Table 1: Vertical protection levels in meters for three choices of per-sample $P_{FA}$.

|  | Min. | Max. | Mean |
|---|---|---|---|
| $P_{FA}^{white}$ | 4.96 | 19.62 | 8.60 |
| $P_{FA}^{corr,common}$ | 4.83 | 18.85 | 8.32 |
| $P_{FA}^{cond}$ | 4.89 | 19.17 | 8.44 |

The table reveals that the maximum VPL computed on the basis of $P_{FA}^{corr,common}$ is about 30 $cm$ lower than it should be, i.e., lower than VPL computed on the basis of compensated conditional $P_{FA}^{cond}$.

The slightly optimistic protection levels stemming from $P_{FA}^{corr,common}$ impact also the availability of the service (it is more optimistic). For example, based on the performed simulations, VPL with 99% availability (for $P_{FA}^{corr,common}$) is

$$VPL_{99} = 11.62\ m,$$

i.e., 99% of all observed VPLs is less than or equal to $VPL_{99}$. However, using the value $VPL_{99}$ in availability analysis of the results for true $P_{FA}^{cond}$ leads to the true availability of 98.61%, which is about 0.4% less than it was expected.

## VI. CONCLUDING REMARKS

The paper dealt with the allocation of the probability of false alert within the advanced receiver integrity monitoring method. The method for correct computation of the per-sample probability of false alert respecting the time-correlation of the underlying test statistics (separation of the sub-solution) was proposed. Because of difficulties in the analytical solution of the proposed relations, the Monte-Carlo based numerical algorithm for the conditional per-sample probability was detailed. The algorithm was illustrated using numerical examples.

## APPENDIX A: IMPACT OF TIME-CORRELATION ON PROBABILITY OF FALSE ALERT

The aim of this appendix is to provide a background and discussion for assessment of the impact of the time correlated pseudorange measurement noise on the probability of false alert. The appendix is split into three parts dealing with the analytical solution, numerical solution, and numerical illustration.

### A. Analytical solution

**Time-uncorrelated random variable:** Let, for the sake of simplicity, a random process $x_k$, described as

$$x_k = w_k, \forall k,$$

be firstly supposed, where $w_k$ is time-uncorrelated (white) random variable with Gaussian probability density function (pdf), $w_k \sim p_{W_k}(w_k) = N(0, Q)$.

Having the specified probability $P_{OUT,k}$, then probability $P_{IN,k}$ and respective bound $q$ (alternatively quantile) can be computed from

$$P_{OUT,k} = 1 - P_{IN,k},$$
$$P_{IN,k} = \int_{-q}^{q} p_{X_k}(x_k) dx_k,$$

where $p_{X_k}(x_k) = p_{W_k}(x_k)$. If both probabilities $P_{IN,k}$ and $P_{OUT,k}$ are constant for all time instants, then $q$ is constant for all time instants as well.

Within this set-up $P_{OUT}$ can be understood as the probability $P_{FA}^{white}$, variable $x$ as the test statistics $d_{Vn}$, and the quantile $q$ as the decision threshold $D_{Vn}$.

**Time-correlated random variable:** As opposed to the previous case, let the random variable $x_k$ be modeled as Gauss-Markov (GM) process of the first order

$$x_{k+1} = ax_k + w_k$$

where $a$ is a parameter closely related to process time constant $\tau$ (for sampling time 1 sec, the parameter is $a = e^{-\frac{1}{\tau}}$) [7].

If the initial condition $x_0$ is Gaussian, i.e. $x_0 \sim p_{X_0}(x_0) = N(0, P_x)$, and the driving noise $w_k$ is Gaussian, i.e. $w_k \sim p_{W_k}(w_k) = N(0, Q), \forall k$, and independent of the initial condition, then the $x_k$ is also Gaussian with the pdf

$$p_{X_k}(x_k) = N(0, P_x)$$

with the steady-state variance

$$P_x = \frac{Q}{1 - a^2}$$

for all time instants.

In the following text, the (conditional) probability is analytically expressed for the first two time instants $k = 0, 1$.

- Time $k = 0$

Given probability $P_{OUT,0}$, bound value $q$ is computed, analogously to time-uncorrelated case, from equation

$$P_{OUT,0} = 1 - \int_{-q}^{q} p_{X_0}(x_0) dx_0,$$

which characterizes the probability that sample $x_0$ is outside interval $S_q = \langle -q, q \rangle$. However, preserving the interval $S_q$ also in the next time instant leads to different value of conditional probability $P_{OUT,1}$.

- Time $k = 1$

Considering the GM process, the probability $P_{OUT,1}$ that the sample $x_1$ is outside interval $S_q$ should be lower than $P_{OUT,0}$. The exact probability $P_{OUT,1}$, as a function of $p_{X_1}(x_1)$, can be computed by "propagating $P_{OUT,0}$", or alternatively $P_{IN,0}$, forward in time under the assumption that sample $x_0 \in S_q$. Forward propagation of $P_{IN,0}$, as a function of $p_{X_0}(x_0)$, is more convenient.

Let validity of the condition $x_0 \in S_q$ assumed ("false alert did not occur at time $k = 0$"). It means that $x_0$ is described by the constrained probability $p_{X_0^B}(x_0|x_0 \in S_q)$ defined as

$$p_{X_0}(x_0|x_0 \in S_q) = c_{norm} \times p_{X_0}(x_0)$$
$$= c_{norm} \times \frac{1}{\sqrt{2\pi P_x}} e^{-\frac{x_0^2}{2P_x}},$$

if $x_0 \in S_q$ and $p_{X_0}(x_0|x_0 \in S_q) = 0$ otherwise. The variable $c_{norm}$ is the normalization constant computed according to $c_{norm} = \frac{1}{1 - P_{OUT,0}}$.

Then, it is possible to determine the transformed pdf of the constrained variable multiplied by the constant $a$ from the GM process using a substitution

$$y = ax_0,$$

which results in the conditional pdf[5]

$$p_{Y_0}(y_0|x_0 \in S_q) = p_{X_0^B}\left(\frac{y_0}{a}\right)\frac{1}{a}$$
$$= c_{norm} \times \frac{1}{\sqrt{2\pi a^2 P_x}} e^{-\frac{y_0^2}{2a^2 P_x}}.$$

if $y_0 \in S_{q,y}$, where $S_{q,y} = \langle -q \times a, q \times a \rangle = \langle -q_y, q_y \rangle$, and $p_{Y_0}(y_0|x_0 \in S_q) = 0$ otherwise.

As a last step, the pdf of $x_1$ can be computed using the convolution[6] of $p_{Y_0}(y_0)$ and $p_{W_0}(w_0)$. Conditional density function $p_{X_1}(x_1|x_0 \in S_q)$ is equal to

$$p_{X_1}(x_1|x_0 \in S_q) = \int_{-\infty}^{\infty} p_{Y_0}(y_0) p_{W_1}(x_1 - y_0) dy_0$$
$$= \int_{-q_y}^{q_y} p_{Y_0}(y_0) p_{W_1}(x_1 - y_0) dy_0$$
$$= c_{norm} \frac{1}{\sqrt{2\pi a^2 P_x}\sqrt{2\pi Q}} \int_{-q_y}^{q_y} e^{-\frac{y_0^2}{2a^2 P_x}} e^{-\frac{(x_1 - y_0)^2}{2Q}} dy_0.$$

Solution to the integral leads to the final relation of conditional density function

$$p_{X_1}(x_1|x_0 \in S_q) = \frac{c_{norm}}{2} \frac{1}{\sqrt{2\pi(a^2 P_x + Q)}} e^{-\frac{x_1^2}{2(a^2 P_x + Q)}}$$
$$\times \left[ \text{erf}\left(\sqrt{\frac{1}{2}} \times \frac{q_y a^2 P_x + q_y Q - a^2 P_x x_1}{\sqrt{a^2 P_x Q(a^2 P_x + Q)}}\right) \right.$$
$$+ \text{erf}\left(\sqrt{\frac{1}{2}}\right.$$
$$\left.\left. \times \frac{q_y a^2 P_x + q_y Q + a^2 P_x x_1}{\sqrt{a^2 P_x Q(a^2 P_x + Q)}}\right) \right]$$

where the error function is defined as $\text{erf}(x) = \frac{2}{\sqrt{\pi}} \int_0^x e^{-t^2} dt$ and the term $(a^2 P_x + Q) = P_x$.

---

[5] Having random variable $x$ with pdf $p_X(x)$ and constant $a$, the pdf of their product $y = ax$ is $p(y) = p_X\left(\frac{y}{a}\right) \times \frac{1}{a}$ [8].

[6] Having two random variables $x$ and $y$ with respective pdf's, the pdf of their sum $z = x + y$ is $p_Z(z) = \int_{-\infty}^{\infty} p_X(x) p_Y(z - x) dx$ [8].

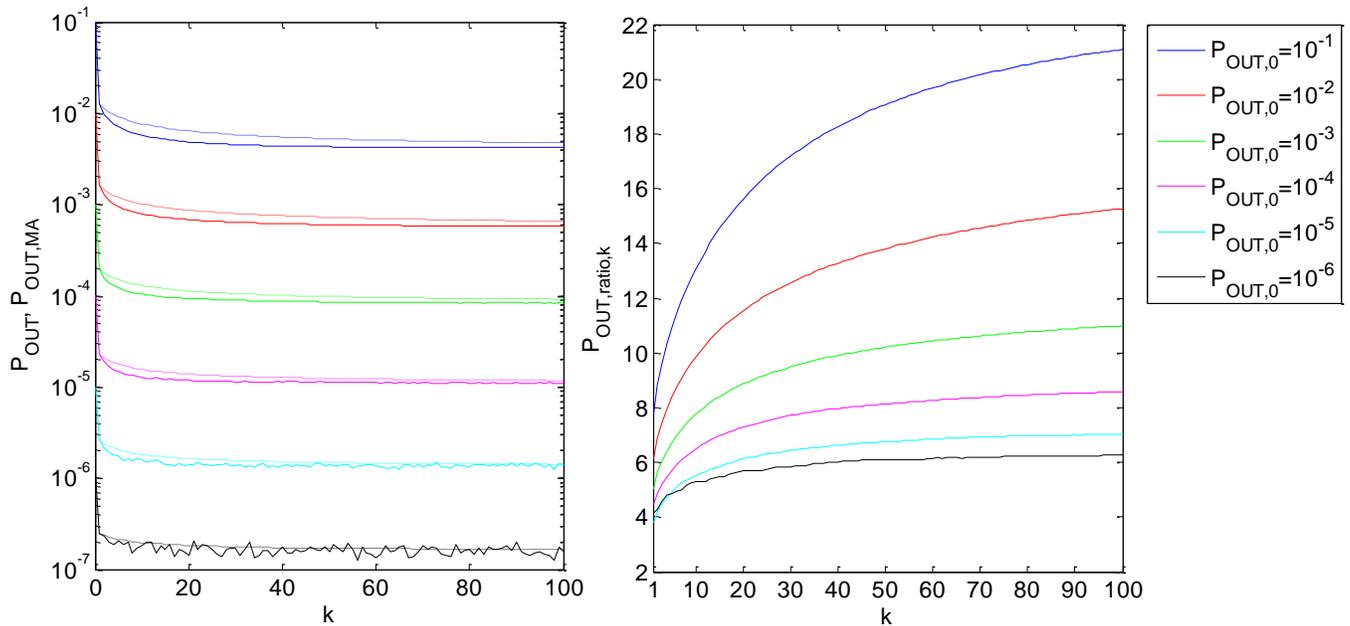

**Figure 4: Time evolution of conditional P$_{OUT}$ (solid line), its MA (dashed line), and the ratio for a=0.99.**

For completeness, the plots of pdf's $p_{X_0}(x_0)$, $p_{X_0^B}(x_0), p_{Y_0}(y_0)$, and $p_{X_1}(x_1)$ are given in Figure 2 for $a = 0.99$, $Q = 0.01, P_x = 0.51$, and $P_{OUT,0} = 0.1$.

**B. Monte-Carlo based algorithm**

From previous part it can be seen that the analytical solution to the conditional probability is not trivial. Therefore, a Monte-Carlo based numerical algorithm is proposed here.

The algorithm computing conditional probability is defined by the following steps.

**Step 1:** Set $k = 0$, define probability $P_{OUT,0}$ and compute the bound $q$.

**Step 2:** Generate samples $\{x_k^{(i)}\}$, $i = 1,2,...,M$, from $p(x_k)$.

**Step 3:** Choose points being in the interval $S_q$, i.e., $-q \leq x_k^{(i)} \leq q$, and denote them as $\{x_{q,k}^{(i)}\}$, $i = 1,2,...,M_q$.

**Step 4:** Transform all the points in bounded set via the GM process equation, i.e., $x_{k+1}^{(i)} = ax_{b,k}^{(i)} + w_k^{(i)}$, where the samples $\{w_k^{(i)}\}$ are generated according to $p_{W_k}(w_k) = N(0, Q)$ and $i = 1,2,...,M_q$.

**Step 5:** Compute number of samples being outside the interval $S_q$, denote it $M_{OUT}$, and compute the conditional probability

$$P_{OUT,k+1} = \frac{M_{OUT}}{M_q}.$$

**Step 6:** Set $k = k + 1$ and go to **Step 3**.

Comparison of the analytically and numerically computed conditional density function $p_{X_1}(x_1|x_0 \in S_q)$ is shown in Figure 3.

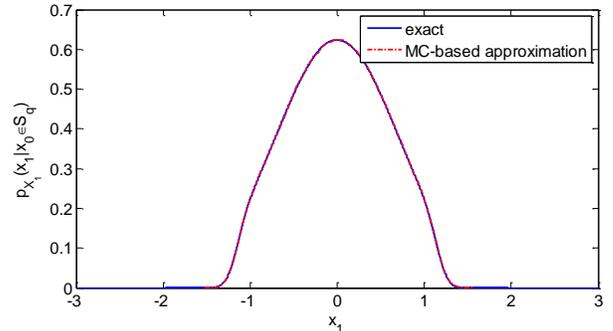

**Figure 3: Comparison of density functions computed analytically and numerically.**

**C. Numerical illustration**

The MC algorithm is illustrated in two examples. In the first one, the conditional probabilities are computed for a GM process with parameters $a = e^{-\frac{1}{100}} = 0.99$, $Q = 0.01$, and thus $P_x = 0.51$. The time behavior of the conditional probability $P_{OUT,k}$, its moving average

$$P_{OUT,MA,k} = \frac{1}{k}\sum_{i=1}^{k} P_{OUT,k}$$

and the ratio

$$P_{OUT,ratio,k} = \frac{P_{OUT,0}}{P_{OUT,MA,k}},$$

are plotted in Figure 4. Note that the ratio is basically the correction coefficient $c_{corr}$ used for compensation of the white-noise based per-sample probability of false alert for time-correlation of the underlying process.

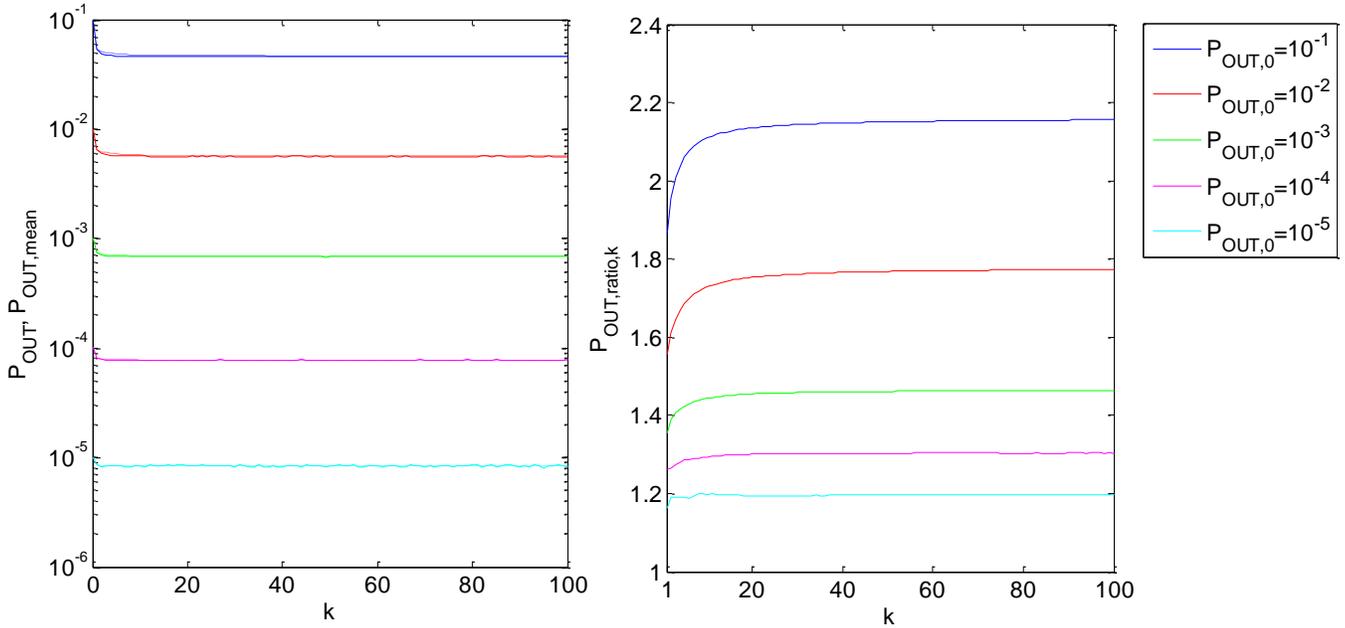

Figure 5: Time evolution of conditional $P_{OUT}$ (solid line), its MA (dashed line), and the ratio for a=0.82.

In the second example, the time constant of the GM process is considered to 5 *sec*, i.e., the process is with $a = e^{-\frac{1}{5}} = 0.82$, and the respective plots are given in Figure 5.

As can be seen, the time evolution of the conditional probability heavily depends on the constant $a$ (determined by the time constant) of the GM process. With the decreasing time constant and thus with decreasing parameter $a$, the GM process is approaching, in terms of its properties, to a white noise, and thus the conditional probability $P_{OUT,k}$, for $k \geq 1$, are "more similar" to the initial one $P_{OUT,0}$. In the limit case for $a = 0$, the probability $P_{OUT,k}$ is constant.

Finally note that the conditional probability computation was performed using the MC-based approach with $5 \times 10^8$ samples which is the reason for seemingly noisy curve for $P_{OUT,0} = 10^{-6}$. In this case the number of samples outside the interval is rather low and computed $P_{OUT,k}$ is a subject of certain random variability. The curve become smoother as the number of samples is increased.

**REFERENCES**

[1] GEAS Phase II of the GNSS Evolutionary Architecture Study, February 2010.
[2] Brenner, M.: "Integrated GPS/Inertial Fault Detection Availability," NAVIGATION: Journal of The Institute of Navigation, vol. 43, no. 2, 1996.
[3] Groves, P. D.: "Principles of GNSS, Inertial, and Multisensor Integrated Navigation Systems," Artech House, 2008.
[4] RTCA DO-229D: Minimum Operational Performance Standards (MOPS) for Global Positioning System / Wide Area Augmentation System Airborne Equipment, December 2006.
[5] Orejas, M., Kana, Z., Dunik, J., Dvorska, J., and Kundak, N.; "Multiconstellation GNSS/INS to Support LPV200 Approaches and Autolanding," Proceedings of the 25th International Technical Meeting of The Satellite Division of the Institute of Navigation (ION GNSS 2012), Nashville, TN, September 2012.
[6] Anderson, B. D. O. Anderson and Moore, J. B.: "Optimal Filtering," New Jersey: Prentice Hall Ins.: Englewood Cliffs, 1979.
[7] Rogers, R. M.. "Applied Mathematics in Integrated Navigation Systems, 3rd edition," American Institute of Aeronautics and Astronautics, 2007.
[8] Grinstead, C. M. and Snell, J. L.: "Introduction to Probability", American Mathematical Society, 1997.

[9] ICAO Annex 10, Aeronautical Telecommunications, Volume I, Radio Navigation Aids, 2006.

[10] Blanch, J., Walter, T., Enge, P., Wallner, S., Fernandez, F. A., Dellago, R., Ioannides, R., Hernandez, I. F., Belabbas, B., Spletter, A., and Rippl, M.: "Critical Elements for a Multi-Constellation Advanced RAIM", NAVIGATION, Journal of The Institute of Navigation, Vol. 60, No. 1, Spring 2013, pp. 53-69.

[11] FAA Global Positioning System Wide Area Augmentation System (WAAS) Performance Standard, 2008.